\documentclass{article}
\usepackage{algorithm}
\usepackage[noend]{algpseudocode}
\usepackage{PRIMEarxiv}
\usepackage{amsmath}
\usepackage{amssymb}
\usepackage{amsthm}
\usepackage{placeins}
\usepackage{xspace}
\usepackage{xcolor}
\usepackage{graphicx}

\usepackage[authoryear,round]{natbib}
\usepackage[hidelinks]{hyperref}

\usepackage[utf8]{inputenc} 
\usepackage[T1]{fontenc}    
\usepackage{hyperref}       
\usepackage{url}            
\usepackage{booktabs}       
\usepackage{amsfonts}       
\usepackage{nicefrac}       
\usepackage{microtype}      
\usepackage{lipsum}
\usepackage{fancyhdr}       
\usepackage{graphicx}       
\graphicspath{{media/}}     

\newcommand{\ourmethod}{Profile-QSurv\xspace}
\newtheorem{theorem}{Theorem}
\newtheorem{proposition}{Proposition}
\newtheorem{corollary}{Corollary}
 
\title{Nonparametric Continuous-Time Neural Hazard Estimation under Heterogeneous Cluster Specific Effects
}

\author{
  Do Gyeong Han \\
  Department of Applied Statistics, Social Sciences, and Humanities \\
  New York University \\
  New York, NY\\
   \And
  Hyungrok Do \\
  Department of Population Health \\
  NYU Grossman School of Medicine \\
  New York, NY\\
}

\begin{document}
\maketitle

\begin{abstract}
In clustered survival data, the hazard depends on patient characteristics and on the cluster, such as the treating center, and differences between clusters can change over follow-up. Existing methods typically describe the cluster effect by a time-constant frailty or by a separate baseline hazard for each cluster. These approaches have two major limitations: i) a time-constant effect does not describe when the differences between clusters arise; ii) separate baseline hazards are imprecise in clusters with few events. We propose \ourmethod, a continuous-time model in which shared neural networks describe how covariates and time relate to the hazard, and each cluster multiplies this hazard by a level and a temporal shape expressed in a shared learned dictionary. The cluster parameters are estimated by penalized profile likelihood without specifying a frailty distribution. We show that the profile estimating equations calibrate the fitted hazard within each cluster, and that the profiled cluster parameters absorb changes in the shared hazard with a shrinkage factor of empirical Bayes form. Through simulations, we compare \ourmethod with pooled, cluster-separate, and frailty models when clusters differ in their temporal patterns and in their sizes. Finally, we apply \ourmethod to a comparison of hospitals for COVID-19 admissions in Sao Paulo, Brazil, where the hospital factors describe the death hazard of each hospital relative to a typical hospital over the stay.
\end{abstract}


\section{Introduction}
\label{sec:introduction}
 
\subsection{Applied Context}
 
Clustered survival data arise when patients are treated within hospitals, transplant centers, or clinical practices. Patients in the same cluster share aspects of their care environment, and both the patient case mix and the event-time distribution can differ across clusters \citep{glidden2004modelling,wintrebert2004centre}. These differences can also change over follow-up. One hospital may have an elevated hazard early after admission and a typical hazard afterward, while another may carry a persistent excess. The instantaneous hazard describes the event rate at each time among patients who remain event-free, and its trajectory shows when cluster differences emerge, persist, or reverse. In contrast, a survival probability at a fixed horizon summarizes the hazard accumulated up to that horizon, and different early and late hazards can lead to the same survival probability.
 
A particularly motivating example that we use in this paper is the comparison of hospitals for patients admitted with coronavirus disease 2019 (COVID-19). In Brazil, in-hospital mortality of COVID-19 varied across regions and over the epidemic, together with the pressure of admissions on hospital resources \citep{ranzani2021characterisation,brizzi2022spatial}. In the city of Sao Paulo, admissions in 2020 were distributed across public, philanthropic, and private hospitals that differed in size, case mix, and organization of care. Excess deaths in the first days after admission may reflect the severity of the patients at presentation and delays in the escalation of care \citep{churpek2016association}. Excess deaths after the second week may instead reflect the care of prolonged critical illness, when the causes of death differ from those early in the course \citep{contou2021causes}. A standardized mortality ratio combines these into a single number. In-hospital mortality also depends on discharge, since a hospital that discharges its patients earlier records fewer of their deaths \citep{drye2012comparison}. This highlights the need for hospital comparisons that describe when in the stay the differences arise, after adjustment for the characteristics of the patients.
 
\subsection{Existing Approaches}
 
Statistical models for clustered survival data treat cluster effects in two broad ways. Shared frailty models describe the cluster effect as a random multiplier of a common hazard, and their estimation is closely connected to penalized likelihood \citep{therneau2003penalized,ripatti2000estimation}. Stratified Cox models give each cluster its own baseline hazard, which is eliminated from the partial likelihood and estimated afterward by Breslow-type estimators \citep{cox1972regression,kalbfleisch2002statistical}. The same two views appear in provider profiling. Hierarchical models shrink the center effects toward the overall mean \citep{normand1997statistical}, and fixed-effects profiling estimates the ratio of observed to expected events of each center without shrinkage \citep{kalbfleisch2013monitoring,he2013evaluating,varewyck2014shrinkage}. Cluster effects that vary over time have been modeled through time-dependent frailties \citep{wintrebert2004centre} and, in time-dynamic profiling, through a fixed time-varying effect for each facility estimated from the data of that facility \citep{estes2018time}. Stratified Cox models have also been extended to covariate effects that vary over time \citep{he2022stratified}. In these models, patient characteristics enter through a linear predictor.
 
Flexible hazard regression relaxes this structure through splines \citep{royston2002flexible}, reduced-rank time-varying effects \citep{perperoglou2013reduced}, and neural networks \citep{katzman2018deepsurv,kvamme2019time,gensheimer2019scalable,tang2022soden,han2022survival}. QSurv parameterizes the instantaneous hazard as a neural function of covariates and time, with Gauss-Legendre quadrature for the cumulative hazard \citep{lee2026qsurv}. For clustered data, neural frailty models combine a neural hazard with a cluster frailty \citep{lee2023deep,zhou2025neural} or an individual frailty \citep{wu2023neural}, and stratified neural Cox models share a neural risk function across strata with separate baselines \citep{kuruc2022stratified}. A neural network has also been used for risk adjustment in the profiling of dialysis facilities \citep{wu2023leveraging}.
 
A key feature of nearly all existing approaches is that the effect of a cluster is either a time-constant multiplier or a function of time estimated from the data of the cluster alone, such as a separate baseline hazard. This has two main limitations: i) a time-constant effect does not describe when in follow-up the differences between clusters arise; ii) a function estimated from one cluster uses no information from the other clusters, so the hazard trajectory of a cluster with few events is imprecise. The approaches with time-varying cluster effects also describe the patient characteristics by a linear predictor. This leaves a clear gap: a continuous-time method that learns nonlinear, time-varying patient associations from all clusters, estimates the absolute hazard trajectory of each cluster, and shares temporal structure across clusters so that clusters with few events remain estimable.
 
In this paper, we propose \ourmethod, a profiled continuous-time neural hazard model for clustered survival data. Building on QSurv \citep{lee2026qsurv}, shared neural networks learn a patient-risk function and a temporal dictionary from all clusters, and each cluster multiplies the shared hazard by a level and a temporal shape expressed in the dictionary. We estimate the model by penalized profile likelihood \citep{murphy2000profile}. The level is profiled in closed form as the ratio of observed to expected events, as in fixed-effects profiling, and the shape is shrunk as a Gaussian random effect \citep{therneau2003penalized}. We show that the profile estimating equations calibrate the fitted hazard within each cluster. We also show that a change in the shared hazard reaches the fitted hazard of a cluster after shrinkage by a factor of empirical Bayes form \citep{ver1996parametric}.
 
The rest of this paper is organized as follows. In Section \ref{sec:model}, we present the proposed model. Section \ref{sec:estimation} describes the penalized profile likelihood and the training of the shared networks. In Section \ref{sec:theory}, we present the calibration and response properties of the profile estimator. We illustrate the performance of \ourmethod in simulation studies in Section \ref{sec:simulation}, and apply it to the motivating example of COVID-19 admissions in Sao Paulo in Section \ref{sec:application}. Finally, we summarize and discuss our findings in Section \ref{sec:discussion}.
 
\section{Problem Formulation}
\label{sec:model}
 
Consider $K$ clusters with $n_k$ patients in cluster $k$ and $n=\sum_{k=1}^{K}n_k$. For patient $i$ in cluster $k$, we observe $(x_{ki},o_{ki},\delta_{ki})$, where $x_{ki}$ denotes the baseline covariates, $o_{ki}$ the earlier of the event and censoring times, measured from cohort entry, and $\delta_{ki}$ the indicator that the event is observed. Throughout the paper, cluster memberships are known and observed, and the clusters are treated as fixed. We assume that patient records are independent given the cluster environments, and that event and censoring times are independent given covariates and cluster. The targets are the cluster-specific hazards $\lambda_k(t|x)$ for $0\leq t\leq\tau$, where $\tau$ is the end of follow-up, on the covariate and follow-up region observed in each cluster.
 
If we ignore cluster memberships, a single hazard can be estimated from all patients, and every cluster receives the same hazard trajectory for patients with the same covariates. Estimating $\lambda_k$ in each cluster separately allows any trajectory but uses only the events of that cluster. If the $K$ clusters are distinct but related, one might wish to borrow strength across the $K$ clusters to estimate the patient-risk association and the temporal pattern, while retaining the differences between clusters. Therefore, instead of estimating one hazard for all clusters or a separate hazard for each cluster, we consider
\begin{equation}
\lambda_k(t|x)=c_k(t)h_\theta(t,x), c_k(t)=a_k\exp\{w_k^{T}\phi_\eta(t)\},
\label{eq:hazard-model}
\end{equation}
herein referred to as \ourmethod. Here $h_\theta(t,x)>0$ is a shared patient-risk network, $\phi_\eta(t)\in\mathbb{R}^{q}$ is a shared temporal network, $a_k>0$ is the level of cluster $k$, and $w_k\in\mathbb{R}^{q}$ contains its shape coefficients. The patient-risk network receives covariates and time as a joint input, as in the concatenation variant of QSurv \citep{lee2026qsurv}, and uses a softplus output for positivity. The temporal network receives only time and supplies the same $q$ features to every cluster.
 
The model separates two kinds of comparison. Within a cluster, $\lambda_k(t|x)/\lambda_k(t|x')=h_\theta(t,x)/h_\theta(t,x')$, which can vary over time and is common to all clusters. For the same covariates in two clusters,
\begin{equation}
\frac{\lambda_k(t|x)}{\lambda_{k'}(t|x)}=\frac{a_k}{a_{k'}}\exp\{(w_k-w_{k'})^{T}\phi_\eta(t)\},
\label{eq:cluster-ratio}
\end{equation}
which depends only on the cluster parameters and the dictionary. Thus, the hazard ratio between two clusters is the same for all patients at each time, and it can change over follow-up and cross one. Each temporal feature is centered and scaled with respect to a uniform reference distribution on $[0,\tau]$, so that time-constant variation is assigned to the levels and the shape coefficients share a common scale for the penalty. A common function of time can move between $h_\theta$ and the cluster factors without changing any $\lambda_k$, so we report the factors relative to a reference cluster. For example, with $c_0(t)=\exp\{K^{-1}\sum_{k=1}^{K}\log c_k(t)\}$, the ratio $c_k(t)/c_0(t)$ is the hazard ratio of cluster $k$ relative to a cluster whose factor is the geometric mean of the cluster factors. Section \ref{sec:application} uses the pointwise median in place of the geometric mean.
 
\begin{figure}[t]
\centering
\includegraphics[width=0.95\linewidth]{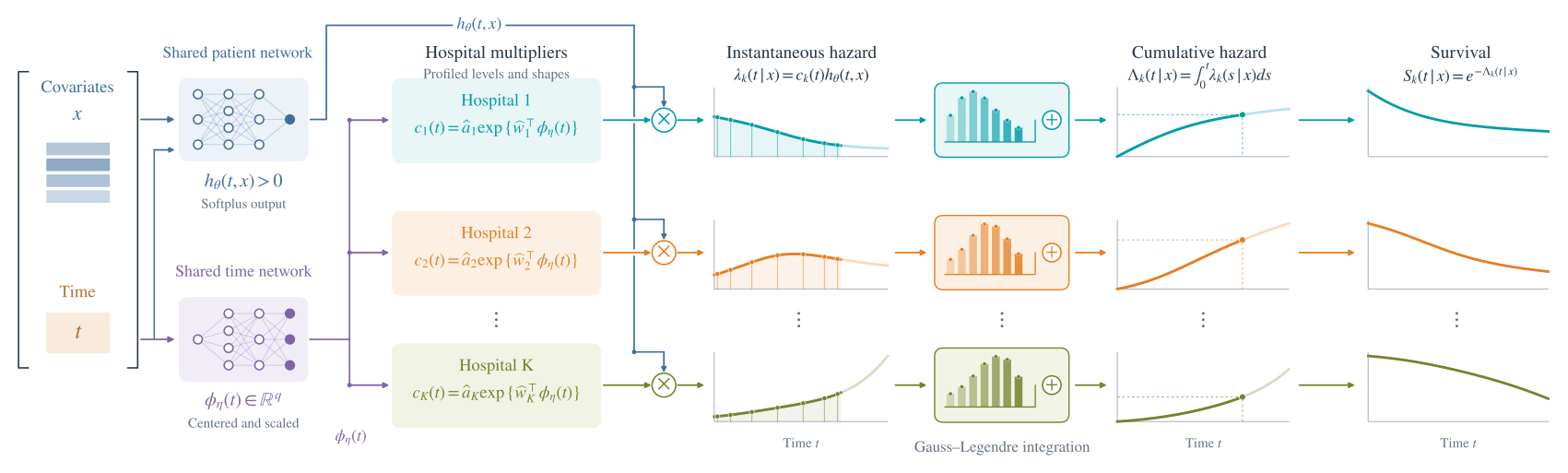}
\caption{Structure of \ourmethod. The shared networks $h_\theta$ and $\phi_\eta$ are learned from all clusters, and the level $a_k$ and shape coefficients $w_k$ of each cluster are profiled from its own events and exposure.}
\label{fig:concept}
\end{figure}

\ourmethod borrows strength across clusters through three components. First, the patient-risk network is learned from the events of all clusters, so nonlinear and time-varying patient associations are estimated from the whole cohort. Second, the temporal dictionary is shared, so each cluster describes its departure from the shared hazard with $q$ coefficients instead of a separate baseline hazard. Similar structures appear in reduced-rank hazard regression \citep{perperoglou2013reduced} and in frailties whose logarithm is a sum of specified functions of time with random coefficients \citep{enki2014time}. Third, the shape coefficients are penalized and the levels are not. As a result, the temporal departure of a cluster with few events is shrunk toward the shared temporal pattern, while its overall event rate stays at its own observed level. Figure \ref{fig:concept} summarizes the structure of the model.
 
\section{Penalized Profile Likelihood Estimation}
\label{sec:estimation}
 
In this section, we introduce the penalized profile likelihood for \ourmethod, in which the parameters of each cluster are profiled out for fixed shared networks, and the shared networks are trained on the resulting profile likelihood.
 
\subsection{Profiling the Cluster Parameters}
 
Ignoring terms that involve only the censoring distribution, the log-likelihood of model \eqref{eq:hazard-model} is
\begin{equation}
\ell(\theta,\eta,a,w)=\sum_{k=1}^{K}\sum_{i=1}^{n_k}\left\{\delta_{ki}\log\lambda_k(o_{ki}|x_{ki})-\int_0^{o_{ki}}\lambda_k(t|x_{ki})dt\right\},
\label{eq:continuous-likelihood}
\end{equation}
and we estimate the parameters by maximizing the penalized criterion
\begin{equation}
\ell_\rho(\theta,\eta,a,w)=\ell(\theta,\eta,a,w)-\frac{\rho}{2}\sum_{k=1}^{K}\|w_k\|^2, \rho>0.
\label{eq:penalized-likelihood}
\end{equation}
Up to a constant, the penalty is the log density of independent Gaussian shape coefficients $w_k\sim N(0,\rho^{-1}I_q)$. Thus, $\rho^{-1}$ plays the role of the variance of the cluster temporal departures, as in the penalized likelihood formulation of Gaussian random effects on the log-hazard scale \citep{therneau2003penalized,ripatti2000estimation}. The levels are unpenalized fixed effects.
 
For fixed networks $(\theta,\eta)$, let $D_k=\sum_{i=1}^{n_k}\delta_{ki}$ and $M_k=\sum_{i=1}^{n_k}\delta_{ki}\phi_\eta(o_{ki})$, and let
\begin{equation}
A_k(w)=\sum_{i=1}^{n_k}\int_0^{o_{ki}}h_\theta(t,x_{ki})\exp\{w^{T}\phi_\eta(t)\}dt
\label{eq:cluster-exposure}
\end{equation}
be the expected number of events in cluster $k$ when its level equals one. We call the distribution over the patients of cluster $k$ and their follow-up times with density $h_\theta(t,x_{ki})\exp\{w^{T}\phi_\eta(t)\}/A_k(w)$ at patient $i$ and time $t\leq o_{ki}$ the exposure distribution of cluster $k$ at $w$, and write $\mathbb{E}_{k,w}$ and $\text{\normalfont Cov}_{k,w}$ for the mean and covariance under it. The terms of $\ell_\rho$ that involve $(a_k,w_k)$ are $D_k\log a_k+w_k^{T}M_k-a_kA_k(w_k)-\rho\|w_k\|^2/2$, and for $D_k>0$, maximizing over the level gives
\begin{equation}
\hat{a}_k(w)=\frac{D_k}{A_k(w)},
\label{eq:profile-level}
\end{equation}
the ratio of the observed events of cluster $k$ to the events expected at level one. This ratio is the counterpart of the standardized event ratio of fixed-effects provider profiling \citep{kalbfleisch2013monitoring,he2013evaluating}. Substituting $\hat{a}_k(w)$, the profile log-likelihood of cluster $k$ is
\begin{equation}
\sum_{i=1}^{n_k}\delta_{ki}\log\frac{h_\theta(o_{ki},x_{ki})\exp\{w^{T}\phi_\eta(o_{ki})\}}{A_k(w)}-\frac{\rho}{2}\|w\|^2+D_k(\log D_k-1).
\label{eq:profile-loglik}
\end{equation}
The ratio inside the logarithm is the density of the exposure distribution at the observed event. Profiling the level therefore replaces the absolute event rate of the cluster by its observed event count. What remains is the log-likelihood of the $D_k$ events under the exposure distribution, which describes how the events are allocated over patients and follow-up time. The stratified Cox model has an analogous structure with a nonparametric stratum baseline, where profiling the baseline leaves the partial likelihood of allocating each event within its risk set \citep{andersen2012statistical,kalbfleisch2002statistical}.
 
The shape coefficients $\hat{w}_k$ maximize \eqref{eq:profile-loglik}, whose negative Hessian in $w$ is
\begin{equation}
D_k\text{\normalfont Cov}_{k,w}(\phi_\eta)+\rho I_q.
\label{eq:profile-hessian}
\end{equation}
Since $\rho>0$, the profile log-likelihood is strongly concave in $w$ and has a unique maximizer even when the learned features are linearly dependent. The clusters are solved separately by damped Newton updates, and the fitted level is $\hat{a}_k=D_k/A_k(\hat{w}_k)$. If $D_k=0$, the level optimum lies at the boundary $a_k=0$ and the shape penalty gives $w_k=0$. The boundary contribution of such a cluster to the profile objective is set to zero, so the cluster contributes no likelihood information to the shared networks. The comparisons of cluster factors, including the reference factor $c_0(t)$ of Section \ref{sec:model}, are made among the clusters with events. Recorded ties are handled by summing the individual continuous-time likelihood contributions at their recorded times.
 
\subsection{Training the Shared Networks}
 
The shared networks minimize the negative penalized profile likelihood normalized by the total sample size,
\begin{equation}
\mathcal{L}(\theta,\eta)=-\frac{1}{n}\ell_\rho\{\theta,\eta,\hat{a}(\theta,\eta),\hat{w}(\theta,\eta)\}.
\label{eq:profile}
\end{equation}
The levels are profiled analytically and the shapes through a $q$-dimensional convex problem, so the nonconvex part of the optimization involves only the shared networks. At an exact inner optimum, the envelope theorem gives the gradient of $\mathcal{L}$ by differentiating $\ell_\rho$ with the fitted cluster parameters held fixed. Each update profiles the cluster parameters on the full training cohort and updates both networks. The exposure integrals are evaluated by Gauss-Legendre quadrature on each observed follow-up interval, which replaces each integral in $A_k(w)$ by a sum of the integrand at the quadrature nodes with positive weights. The level update and the shape problem keep the same form with this approximation of $A_k$, so profiling is exact for the quadrature criterion up to the inner solver tolerance. The quadrature error is controlled by the temporal smoothness of the hazard, as in QSurv \citep{lee2026qsurv}, and the modeled hazard remains a continuous function of time. At an exact inner optimum, the gradient of the profile objective with respect to the networks includes the dependence of the temporal features and their normalization on $\eta$. Each temporal feature is centered with respect to a uniform reference distribution on $[0,\tau]$ and scaled by its reference standard deviation. The reference moments depend only on the temporal network, so the normalization is common to all clusters and all patient chunks, and the reference standard deviation is floored at a small constant so that nearly constant features keep a bounded scale. Algorithm \ref{alg:profile} summarizes the training of \ourmethod.
 
\begin{algorithm}[t]
\caption{Training \ourmethod}
\label{alg:profile}
\begin{algorithmic}[1]
\Require Training data $\{(x_{ki},o_{ki},\delta_{ki})\}$, penalty $\rho$, dictionary dimension $q$, Gauss-Legendre nodes and weights
\State Initialize the network parameters $(\theta,\eta)$
\Repeat
\For{each cluster $k$ with $D_k>0$}
\State Maximize \eqref{eq:profile-loglik} over $w$ by damped Newton updates to obtain $\hat{w}_k$
\State Set $\hat{a}_k=D_k/A_k(\hat{w}_k)$
\EndFor
\State Evaluate the validation criterion at the current networks and profiles
\State Compute the gradient of the profile objective with $\{(\hat{a}_k,\hat{w}_k)\}$ held fixed, accumulating over patient chunks
\State Update $(\theta,\eta)$ by a gradient step
\Until{the validation criterion stops improving}
\State \Return $(\hat{\theta},\hat{\eta})$ and $\{(\hat{a}_k,\hat{w}_k)\}_{k=1}^{K}$ profiled at the final networks
\end{algorithmic}
\end{algorithm}
 
The penalty $\rho$ and the dictionary dimension $q$ determine how much temporal heterogeneity a cluster can express. As $\rho\rightarrow\infty$, all shapes shrink to zero and the model becomes a fixed-effects model with a time-constant cluster multiplier, $\lambda_k(t|x)=a_kh_\theta(t,x)$. As $\rho\rightarrow0$, the shapes are unpenalized, and the model becomes a stratified model whose log cluster factors lie in the span of the dictionary and a constant. A finite solution may then not exist for a cluster with few events. We use $q=3$ in this paper. We select the penalty together with the architecture of the patient-risk network and the number of training epochs by the held-out log-likelihood, with the cluster parameters profiled on the training data, as in Sections \ref{sec:simulation} and \ref{sec:application}.
 
\section{Properties of the Profile Estimator}
\label{sec:theory}
 
We now present properties of the profile estimator that describe how each cluster borrows information from the shared networks. Fix a cluster $k$ with $D_k>0$, suppress the network subscripts, and let $(\hat{a}_k,\hat{w}_k)$ be its exact penalized profile solution at shared functions $(h,\phi)$. We write $\hat{\lambda}_k(t|x)=\hat{a}_k\exp\{\hat{w}_k^{T}\phi(t)\}h(t,x)$ for the fitted hazard, and $\mathbb{E}_k$, $\text{\normalfont Var}_k$, and $\text{\normalfont Cov}_k$ for moments under the exposure distribution at $\hat{w}_k$, which we call the fitted exposure distribution. The results condition on the observed cohort and hold for any parameterization of the networks. With fixed positive quadrature weights, they also hold exactly for the quadrature criterion.
 
\subsection{Calibration Within Clusters}
 
Since $\hat{a}_kA_k(\hat{w}_k)=D_k$, the fitted exposure distribution has density $\hat{\lambda}_k(t|x_{ki})/D_k$ at patient $i$ and time $t\leq o_{ki}$. That is, it is the distribution of the expected events of the cluster over patients and follow-up time.
 
\begin{proposition}[Profile calibration]
\label{prop:calibration}
Let $D_k>0$. The penalized profile solution of cluster $k$ satisfies
\begin{equation}
\sum_{i=1}^{n_k}\int_0^{o_{ki}}\hat{\lambda}_k(t|x_{ki})dt=D_k, \sum_{i=1}^{n_k}\left\{\delta_{ki}\phi(o_{ki})-\int_0^{o_{ki}}\phi(t)\hat{\lambda}_k(t|x_{ki})dt\right\}=\rho\hat{w}_k.
\label{eq:calibration}
\end{equation}
\end{proposition}
\begin{proof}
The derivative of the penalized log-likelihood in $a_k$ vanishes at the profile solution, which gives $\hat{a}_kA_k(\hat{w}_k)=D_k$. Since $\hat{a}_kA_k(\hat{w}_k)=\sum_{i=1}^{n_k}\int_0^{o_{ki}}\hat{\lambda}_k(t|x_{ki})dt$, this is the first identity in \eqref{eq:calibration}. The derivative in $w_k$ gives $\hat{a}_k\nabla_wA_k(\hat{w}_k)-M_k+\rho\hat{w}_k=0$, where
\begin{equation}
\hat{a}_k\nabla_wA_k(\hat{w}_k)=\sum_{i=1}^{n_k}\int_0^{o_{ki}}\phi(t)\hat{\lambda}_k(t|x_{ki})dt,
\end{equation}
which gives the second identity.
\end{proof}
 
The first identity states that the fitted cluster hazard reproduces the observed number of events, that is, the martingale residuals of the cluster sum to zero \citep{therneau1990martingale}. Every cluster is therefore calibrated in the large on its own training data, for any event count and any shared networks. The second identity states that observed and expected events also agree along the temporal features, up to the imbalance $\rho\hat{w}_k$ allowed by the penalty.
 
The shape coefficients are bounded by the imbalance of a time-constant multiplier with the same networks,
\begin{equation}
\|\hat{w}_k\|_2\leq\frac{1}{\rho}\|M_k-D_k\mathbb{E}_{k,0}[\phi]\|_2,
\label{eq:web-shape-bound}
\end{equation}
where $M_k-D_k\mathbb{E}_{k,0}[\phi]$ is the difference between the observed and expected events along the temporal features under the time-constant multiplier $D_k/A_k(0)$. Since $\log A_k(w)$ is convex with gradient $\mathbb{E}_{k,w}[\phi]$, we have $\hat{w}_k^{T}\{\mathbb{E}_k[\phi]-\mathbb{E}_{k,0}[\phi]\}\geq0$, and the second identity in \eqref{eq:calibration}, $\rho\hat{w}_k=M_k-D_k\mathbb{E}_k[\phi]$, gives
\begin{equation}
\rho\|\hat{w}_k\|_2^2=\hat{w}_k^{T}\{M_k-D_k\mathbb{E}_k[\phi]\}\leq\hat{w}_k^{T}\{M_k-D_k\mathbb{E}_{k,0}[\phi]\}\leq\|\hat{w}_k\|_2\|M_k-D_k\mathbb{E}_{k,0}[\phi]\|_2,
\end{equation}
which proves \eqref{eq:web-shape-bound}. Since the solution is unique, $\hat{w}_k=0$ if and only if $M_k=D_k\mathbb{E}_{k,0}[\phi]$, that is, if and only if the time-constant multiplier balances the observed and expected events along every temporal feature.
 
\subsection{Response to Changes in the Shared Networks}
 
Consider continuously differentiable perturbations of both shared functions, $\log h_\varepsilon(t,x)=\log h(t,x)+\varepsilon r(t,x)+o(\varepsilon)$ and $\phi_\varepsilon(t)=\phi(t)+\varepsilon\psi(t)+o(\varepsilon)$. For each $\varepsilon$, both cluster parameters are re-estimated with the same cohort and penalty, and $\hat{\lambda}_{k,\varepsilon}$ denotes the resulting fitted hazard. We write $\partial_\varepsilon$ for the derivative in $\varepsilon$ at $\varepsilon=0$. The perturbation $\psi$ includes the effect of the feature normalization wherever it is differentiable. At fixed cluster parameters, the derivative of the log of the unnormalized exposure density is $e_k(t,x)=r(t,x)+\hat{w}_k^{T}\psi(t)$, and
\begin{equation}
\Delta_k(\psi)=\sum_{i=1}^{n_k}\left\{\delta_{ki}\psi(o_{ki})-\int_0^{o_{ki}}\psi(t)\hat{\lambda}_k(t|x_{ki})dt\right\}
\label{eq:theory-perturbation}
\end{equation}
is the difference between the observed and expected events of the cluster along $\psi$, as in the second identity of \eqref{eq:calibration}. At $\hat{w}_k$, the matrix in \eqref{eq:profile-hessian} is $D_k\text{\normalfont Cov}_k(\phi)+\rho I_q$, the penalized profile information for the shape coefficients.
 
\begin{theorem}[Response of a profiled cluster hazard]
\label{thm:response}
Suppose the exposure is positive and finite in a neighborhood of $\varepsilon=0$, the temporal features and perturbations are bounded, and differentiation under the exposure integral is valid. Assume that the exposure moments and their first derivatives in $(w,\varepsilon)$ are continuous near $(\hat{w}_k,0)$. Then the profile solution is locally differentiable, and
\begin{equation}
\begin{split}
\partial_\varepsilon\log\hat{\lambda}_{k,\varepsilon}(t|x)&=e_k(t,x)-\mathbb{E}_k[e_k]\\
&+\{\phi(t)-\mathbb{E}_k[\phi]\}^{T}\{D_k\text{\normalfont Cov}_k(\phi)+\rho I_q\}^{-1}\{\Delta_k(\psi)-D_k\text{\normalfont Cov}_k(\phi,e_k)\}.
\end{split}
\label{eq:theory-response}
\end{equation}
\end{theorem}
\begin{proof}
For clarity, suppress the cluster index, so that $D=D_k$, $e=e_k$, $\Delta(\psi)=\Delta_k(\psi)$, and $\mathbb{E}$ and $\text{\normalfont Cov}$ are moments under the fitted exposure distribution. Let $\hat{w}_\varepsilon$ and $\hat{a}_\varepsilon$ denote the profile solution at $\varepsilon$, and $\mathbb{E}_{\varepsilon,w}$ the expectation under the exposure distribution whose density at patient $i$ and time $t\leq o_i$ is proportional to $h_\varepsilon(t,x_i)\exp\{w^{T}\phi_\varepsilon(t)\}$. By the second identity of Proposition \ref{prop:calibration}, $\hat{w}_\varepsilon$ solves
\begin{equation}
D\mathbb{E}_{\varepsilon,w}[\phi_\varepsilon]-\sum_{i}\delta_i\phi_\varepsilon(o_i)+\rho w=0.
\label{eq:proof-score}
\end{equation}
At $(\hat{w},0)$, the derivative of the left side in $w$ is $D\text{\normalfont Cov}(\phi)+\rho I_q$, which is positive definite. Bounded features and the positive quadratic penalty make the profile log-likelihood strongly concave, so its maximizer is unique. The implicit function theorem therefore gives a differentiable local solution $\hat{w}_\varepsilon$ under the stated smoothness assumptions, also when $\text{\normalfont Cov}(\phi)$ is singular.
 
At fixed $w=\hat{w}$, the log derivative of the unnormalized exposure is $e=r+\hat{w}^{T}\psi$. Differentiating a normalized expectation gives
\begin{equation}
\partial_\varepsilon\mathbb{E}_{\varepsilon,\hat{w}}[\phi_\varepsilon]=\mathbb{E}[\psi]+\text{\normalfont Cov}(\phi,e).
\end{equation}
Since $\sum_i\delta_i\psi(o_i)-D\mathbb{E}[\psi]=\Delta(\psi)$, differentiating \eqref{eq:proof-score} along the fitted path gives
\begin{equation}
\{D\text{\normalfont Cov}(\phi)+\rho I_q\}\partial_\varepsilon\hat{w}_\varepsilon+D\text{\normalfont Cov}(\phi,e)-\Delta(\psi)=0,
\end{equation}
and hence
\begin{equation}
\partial_\varepsilon\hat{w}_\varepsilon=\{D\text{\normalfont Cov}(\phi)+\rho I_q\}^{-1}\{\Delta(\psi)-D\text{\normalfont Cov}(\phi,e)\}.
\label{eq:proof-wdot}
\end{equation}
The level estimate is $\hat{a}_\varepsilon=D/A_\varepsilon(\hat{w}_\varepsilon)$, so that
\begin{equation}
\partial_\varepsilon\log\hat{a}_\varepsilon=-\mathbb{E}[e+\phi^{T}\partial_\varepsilon\hat{w}_\varepsilon]=-\mathbb{E}[e]-\mathbb{E}[\phi]^{T}\partial_\varepsilon\hat{w}_\varepsilon.
\end{equation}
Since $\log\hat{\lambda}_\varepsilon=\log h_\varepsilon+\log\hat{a}_\varepsilon+\hat{w}_\varepsilon^{T}\phi_\varepsilon$,
\begin{equation}
\partial_\varepsilon\log\hat{\lambda}_\varepsilon=e-\mathbb{E}[e]+\{\phi-\mathbb{E}[\phi]\}^{T}\partial_\varepsilon\hat{w}_\varepsilon.
\end{equation}
Substituting \eqref{eq:proof-wdot} proves \eqref{eq:theory-response}.
\end{proof}
 
The response removes two parts of $e_k$. The level absorbs the average of $e_k$ under the fitted exposure, and the shape absorbs the part of the centered change that the temporal features predict in a ridge regression weighted by the fitted exposure. The term in $\Delta_k(\psi)$ is the response of the shape to the imbalance between observed and expected events along the change in the features. It vanishes when the fitted cluster model is calibrated along $\psi$. With a fixed dictionary, the response has the following spectral form.
 
\begin{corollary}[Shrinkage of shared changes with a fixed dictionary]
\label{cor:attenuation}
Let $\psi=0$, let $\sigma_j$ be the positive eigenvalues of $\text{\normalfont Cov}_k(\phi)$ with orthonormal eigenvectors $v_j$, and let $u_j(t)=v_j^{T}\{\phi(t)-\mathbb{E}_k[\phi]\}/\sigma_j^{1/2}$ be the standardized principal components of the temporal features under the fitted exposure distribution. Then
\begin{equation}
\partial_\varepsilon\log\hat{\lambda}_{k,\varepsilon}(t|x)=r(t,x)-\mathbb{E}_k[r]-\sum_{j}\frac{D_k\sigma_j}{D_k\sigma_j+\rho}\text{\normalfont Cov}_k(r,u_j)u_j(t),
\label{eq:theory-attenuation}
\end{equation}
and consequently $\mathbb{E}_k[\{\partial_\varepsilon\log\hat{\lambda}_{k,\varepsilon}\}^2]\leq\text{\normalfont Var}_k(r)$. Moreover, if $h$ is multiplied by a positive constant and $\phi$ is unchanged, the fitted hazard of every cluster is unchanged.
\end{corollary}
\begin{proof}
With $\psi=0$, we have $e_k=r$ and $\Delta_k(\psi)=0$, so that \eqref{eq:theory-response} becomes
\begin{equation}
\partial_\varepsilon\log\hat{\lambda}_{k,\varepsilon}(t|x)=r(t,x)-\mathbb{E}_k[r]-\{\phi(t)-\mathbb{E}_k[\phi]\}^{T}\{D_k\text{\normalfont Cov}_k(\phi)+\rho I_q\}^{-1}D_k\text{\normalfont Cov}_k(\phi,r).
\end{equation}
We expand the inverse matrix in the eigenvectors of $\text{\normalfont Cov}_k(\phi)$. An eigenvector $v$ with zero eigenvalue contributes nothing, since $v^{T}\phi$ has zero variance under the fitted exposure distribution and hence $v^{T}\text{\normalfont Cov}_k(\phi,r)=0$. Since $v_j^{T}\{\phi(t)-\mathbb{E}_k[\phi]\}=\sigma_j^{1/2}u_j(t)$ and $v_j^{T}\text{\normalfont Cov}_k(\phi,r)=\sigma_j^{1/2}\text{\normalfont Cov}_k(r,u_j)$, the eigenvector $v_j$ contributes $D_k\sigma_j\text{\normalfont Cov}_k(r,u_j)u_j(t)/(D_k\sigma_j+\rho)$, which proves \eqref{eq:theory-attenuation}. Under the fitted exposure distribution, the $u_j$ have mean zero and unit variance and are uncorrelated. Thus
\begin{equation}
\mathbb{E}_k[\{\partial_\varepsilon\log\hat{\lambda}_{k,\varepsilon}\}^2]=\text{\normalfont Var}_k(r)-\sum_j\left\{1-\frac{\rho^2}{(D_k\sigma_j+\rho)^2}\right\}\text{\normalfont Cov}_k(r,u_j)^2\leq\text{\normalfont Var}_k(r).
\end{equation}

For the constant rescaling, take $h'=\exp(\kappa)h$ for a constant $\kappa$ with $\phi$ unchanged. Then $A_k(w)$ is multiplied by $\exp(\kappa)$, and \eqref{eq:profile-loglik} is unchanged. Thus $\hat{w}_k$ is unchanged and $\hat{a}_k$ is multiplied by $\exp(-\kappa)$, which preserves the fitted hazard pointwise.
\end{proof}
 
The fraction of a small centered shared change along $u_j$ that remains in the fitted hazard after profiling is $1-D_k\sigma_j/(D_k\sigma_j+\rho)=\rho/(D_k\sigma_j+\rho)$. Here $D_k\sigma_j$ is the profile information for the shape coefficient along $v_j$ after the level is eliminated, and $\rho$ is the precision of the Gaussian prior implied by the penalty. The factor $\rho/(D_k\sigma_j+\rho)$ is the shrinkage weight of the corresponding normal-normal problem \citep{ver1996parametric}. A cluster with many events absorbs shared changes along its well-measured temporal directions and keeps its own temporal pattern. For a cluster with few events, or along a direction with little variation under the exposure of the cluster, the shared change carries over to the fitted hazard, and the cluster follows the temporal pattern learned from all clusters. Profiling leaves within-cluster patient contrasts unchanged, since the cluster factor depends only on time.
 
When the dictionary is also learned, Theorem \ref{thm:response} gives the following bound.
 
\begin{corollary}[Response to dictionary learning]
\label{cor:dictionary}
Under the conditions of Theorem \ref{thm:response},
\begin{equation}
\mathbb{E}_k[\{\partial_\varepsilon\log\hat{\lambda}_{k,\varepsilon}\}^2]^{1/2}\leq\text{\normalfont Var}_k(e_k)^{1/2}+\frac{\|\Delta_k(\psi)\|_2}{2(D_k\rho)^{1/2}}.
\label{eq:web-feature-bound}
\end{equation}
\end{corollary}
\begin{proof}
By the proof of Corollary \ref{cor:attenuation}, the right side of \eqref{eq:theory-response} is the right side of \eqref{eq:theory-attenuation} with $r$ replaced by $e_k$, plus $\{\phi(t)-\mathbb{E}_k[\phi]\}^{T}\{D_k\text{\normalfont Cov}_k(\phi)+\rho I_q\}^{-1}\Delta_k(\psi)$. By the same proof, the first part has second moment at most $\text{\normalfont Var}_k(e_k)$ under the fitted exposure distribution. The second moment of the second part is
\begin{equation}
\Delta_k(\psi)^{T}\{D_k\text{\normalfont Cov}_k(\phi)+\rho I_q\}^{-1}\text{\normalfont Cov}_k(\phi)\{D_k\text{\normalfont Cov}_k(\phi)+\rho I_q\}^{-1}\Delta_k(\psi).
\end{equation}
Since $\sup_{\sigma\geq0}\sigma/(D_k\sigma+\rho)^2=1/(4D_k\rho)$, its square root is at most $\|\Delta_k(\psi)\|_2/\{2(D_k\rho)^{1/2}\}$. The triangle inequality proves \eqref{eq:web-feature-bound}.
\end{proof}
The second term is the norm of the difference between the observed and expected events along the feature change, divided by $2(D_k\rho)^{1/2}$. Beyond its response to $e_k$, the fitted hazard of a cluster therefore changes only through the miscalibration of the cluster along the change in the features. The penalty bounds this change uniformly over the spectrum of $\text{\normalfont Cov}_k(\phi)$. These results hold for any positive event count, including singular feature covariance matrices and zero fitted shape coefficients.

\section{Simulation Study}
\label{sec:simulation}

We performed a series of simulations to evaluate \ourmethod, and compared it with the approaches of a pooled QSurv model, a local QSurv model, a neural gamma frailty model, and a stratified neural Cox model. The pooled model fits one hazard network to all patients without hospital membership, and the local model fits a separate hazard network in each hospital, both with the same architecture as $h_\theta$ \citep{lee2026qsurv}. The frailty model is a proportional hazards model with a neural risk function, a baseline hazard with jumps at the event times, and a gamma frailty for each hospital, as in the deep frailty models for clustered survival data \citep{lee2023deep,zhou2025neural}. The stratified model shares a neural risk function across the hospitals and has a separate baseline hazard for each hospital \citep{kuruc2022stratified}.

\subsection{Simulation Setup}

For Scenarios 1 and 3, we consider five hospitals, H1 to H5. The training samples were simulated as follows. Each patient had five covariates drawn independently from the uniform distribution on $[-1,1]$, with the same distribution in every hospital. The hazard of a patient with covariates $x$ in hospital $k$ was
\begin{equation}
\lambda_k(t|x)=0.015a_k\exp\{f(x)+(0.8x_1+sb_k)(2t/27-1)\}, 0\leq t\leq27,
\label{eq:sim-hazard}
\end{equation}
with a nonlinear function $f$ of the covariates, slopes $(b_1,\cdots,b_5)=(-1.5,-0.75,0,0.75,1.5)$, and a scale $s\in[0,1]$ for the difference in temporal pattern between the hospitals. The shared part $0.015\exp\{f(x)+0.8x_1(2t/27-1)\}$ is nonlinear in the covariates. The hazard ratio between $x_1=1$ and $x_1=-1$ with the other covariates fixed changes from $\exp(-1.6)$ at time zero to $\exp(1.6)$ at day 27. The hospital factor $a_k\exp\{sb_k(2t/27-1)\}$ has a log-linear trend with slope $sb_k$, and its level was set to $a_k=sb_k/\sinh(sb_k)$, with $a_k=1$ for $sb_k=0$. With this choice, every hospital has the same 27-day mortality of 33\% for a patient with $x=0$ at every $s$, so the hospitals differ at $x=0$ only in when their deaths occur. Event times were drawn by inverting the cumulative hazard, and patients without an event by day 27 were censored at day 27.

For Scenarios 2, 4, and 5, we consider a larger number of hospitals, following the simulation design of \citet{wu2023leveraging} for provider profiling. Each patient had three covariates $z=(z_1,z_2,z_3)$, and the hazard of a patient in hospital $k$ was
\begin{equation}
\lambda_k(t|z)=\lambda_0\exp\{\gamma_k+d_{k1}\psi_1(t)+d_{k2}\psi_2(t)+g(z)+0.8z_1(2t/27-1)\}, 0\leq t\leq27,
\label{eq:sim-many}
\end{equation}
where $\psi_1(t)=\exp(-t/2)$ is a departure in the first days, $\psi_2(t)=[1+\exp\{-(t-14)/2\}]^{-1}$ is a departure after the second week, and $g$ is a nonlinear function of the covariates. The levels $\gamma_k$ and the departures $d_{k1}$ and $d_{k2}$ were drawn independently from $N(0,0.4^2)$, and the hospital sizes from a Poisson distribution with mean 100. The covariates were normal with unit variances and pairwise correlations $r$, and each covariate had correlation $r$ with the level $\gamma_k$, as in \citet{kalbfleisch2013monitoring}, so that the case mix of a hospital depends on its level when $r>0$. The baseline rate $\lambda_0$ was set so that 30\% of the patients die by day 27.

The functions of the covariates in the two designs are
\begin{equation}
\begin{aligned}
f(x)&=0.8\sin(\pi x_1)+0.6x_2^2+0.5x_3x_4+0.4x_5,\\
g(z)&=\{z_1+z_2/2-z_3+0.2z_1z_2+0.8|z_2|+0.4\cos(z_1)\sin(z_3)\}/2.
\end{aligned}
\label{eq:sim-functions}
\end{equation}
Each function combines linear effects with nonlinear effects and an interaction between covariates, so that a linear predictor is misspecified in both designs. The function $g$ is the nonlinear function of \citet{wu2023leveraging}, adapted to a hazard as described below. The simulations were conducted under five scenarios, on the temporal pattern of the hospital differences (Scenarios 1 and 2), the information within a hospital and across hospitals (Scenarios 3 and 4), and the case mix of the hospitals (Scenario 5).
\begin{itemize}
\item \textbf{Scenario 1 (Difference in Temporal Pattern):} The scale $s$ ranged from 0 to 1. With $s=0$, the five hospitals have the same hazard and the pooled model is correctly specified. With $s=1$, the hazard ratio of H1 to H5 moves from $e^{3}$ at time zero to $e^{-3}$ at day 27. The sample sizes were set at 1,200 for H1 and H5 and 150 for H2, H3, and H4.
\item \textbf{Scenario 2 (Shape of the Temporal Pattern):} We simulated 25 hospitals with a mean size of 400, each with one episode of excess or deficit $a_k\exp\{-(t-u_k)^2/(2w^2)\}$ in place of the two departures of \eqref{eq:sim-many}, where the day $u_k$ was drawn uniformly on $[0,27]$. The width $w$ was set at 6, 3, 2, 1.5, and 1 days. As $w$ decreases, the share of the variation of the episodes outside the best three functions of time increases from 0.5\% at $w=6$ to 49\% at $w=1$, while \ourmethod uses $q=3$ temporal features.
\item \textbf{Scenario 3 (Hospital Size):} We varied how much a hospital can learn about its own temporal pattern from its own patients. H2, H3, and H4 had between 50 and 1,200 patients each, while H1 and H5 were fixed at 1,200 patients and $s=1$. With 50 patients, a hospital has about a dozen deaths in its training set and its temporal pattern has to be borrowed from the other hospitals, while with 1,200 patients it has about 300 deaths and its temporal pattern can be estimated from the hospital alone.
\item \textbf{Scenario 4 (Number of Hospitals):} We varied how many hospitals contribute to the shared networks. We simulated 100 hospitals from \eqref{eq:sim-many} with $r=0$, and kept the first $K=5$, 10, 25, and 50 of them or all $K=100$.
\item \textbf{Scenario 5 (Case Mix and Hospital Level):} We simulated 100 hospitals from \eqref{eq:sim-many}, with the correlation $r$ between the covariates and the hospital level set at 0, 0.25, 0.5, 0.75, and 0.9. With $r>0$, patients at higher risk are admitted to the hospitals with higher levels, and a model has to separate the case mix from the hospital level.
\end{itemize}
In an additional setting with proportional hospital effects, every hospital has the slope $-0.75$ in place of $sb_k$ in \eqref{eq:sim-hazard}, the hospital factors differ by the constant multipliers $\exp(-0.4)$, $\exp(-0.2)$, 1, $\exp(0.2)$, and $\exp(0.4)$ for H1 to H5, and the sample sizes are those of Scenario 1.

The function $g$ of Scenarios 2, 4, and 5 is half of the nonlinear function of \citet{wu2023leveraging}, with $0.8|z_2|$ in place of their $0.8z_2^2$, since $\exp(0.8z_2^2)$ has an infinite mean under normal covariates. The covariates of a patient in hospital $k$ follow $N\{(r/0.4)\gamma_k\mathbf{1},\Omega-r^2J\}$ with $\Omega=(1-r)I+rJ$, where $J$ is the matrix of ones, the hospital sizes were truncated below at 20, and $\lambda_0=0.0067$. The sizes, levels, and departures of the hospitals were drawn once and kept in every dataset. In Scenario 2, the hospitals have a mean size of 400 and the hazard is $\lambda_0\exp[\gamma_k+a_k\exp\{-(t-u_k)^2/(2w^2)\}+g(z)]$ with $\lambda_0=0.0076$ and $r=0$. The sign of $a_k$ was drawn with probability $1/2$, and $|a_k|$ is common to the hospitals and set so that the root mean square of the centered log hospital factors is 0.25 at every $w$. The reference patients were drawn from $N(0,\Omega)$. For the coverage in Scenario 5, the reference is the pointwise median of the hospital factors, as in Section \ref{sec:application}.

We evaluated the methods against the true hazards on an independent reference population of 512 patients, with every reference patient evaluated in every hospital. The patient-level error is the root mean squared error of the hazard over the reference patients and follow-up. The error of hospital-average hazards uses $\bar{\lambda}_k(t)=512^{-1}\sum_{i=1}^{512}\lambda_k(t|x_i)$, the hazard trajectory of hospital $k$ standardized to the reference population. All hazards were compared as averages over the 54 intervals between 55 equally spaced times on $[0,27]$. Errors are in events per 1,000 patient-days and were averaged over the hospitals. Medians and interquartile ranges (IQRs) of the errors were generated from 20 replicates for each experiment.

We also evaluated the coverage of the pointwise 95\% intervals of Section \ref{sec:application} for the hospital hazard ratios $c_k(t)/c_0(t)$ in each dataset of Scenario 1 with $s=0$ and $s=1$, using the configuration, penalty, and number of epochs selected for \ourmethod. The patients were split into halves within each hospital and event status, the networks were trained on one half and the hospital parameters were profiled on the other, and the point estimates and the 200 parameter draws were averaged over the two assignments of the halves. The intervals were formed for $\log\{c_k(t)/c_0(t)\}$ of Section \ref{sec:model} at days $0,1,\cdots,27$, with the hospital parameters profiled at the selected penalty and, with the same networks, at $\rho=1$. The geometric mean is the reference in the simulations since, with five hospitals, the pointwise median is the factor of one of the hospitals at each time. The coverage of a hospital is the proportion of the 28 days at which its pointwise 95\% interval contains the true value. In Scenario 5, we also evaluated these intervals against their true values.

\subsection{Choice of the Hyperparameters}

Within each hospital, the patients were split into training, validation, and test sets in proportions 0.6, 0.2, and 0.2. In each dataset, ten network configurations were sampled from the 54 combinations of hidden-layer widths of 32, 64, and 128, one to three hidden layers, learning rates of 0.001, 0.003, and 0.01, and weight decay of 0 and $10^{-4}$, and were the same for all methods. For each method, the configuration and the number of epochs, up to 300, were selected by the validation log-likelihood, with the hospitals weighted equally and the epochs of the local model selected in each hospital. \ourmethod used $q=3$ features from a temporal network with one hidden layer of eight units. Each of its configurations was paired with a penalty $\rho$ from $\{1,10,100\}$, with each penalty assigned to three or four configurations at random, so that the penalty was selected together with the configuration and every method had ten fits. The QSurv-based fits used 15 Gauss-Legendre nodes on each of the intervals $[0,2]$, $[2,6]$, and $[6,27]$. The gamma frailty in the neural frailty model has mean one and an estimated variance, and its configuration and number of epochs were selected by the validation partial log-likelihood given the estimated frailties. The Breslow estimators of the baseline hazards of the frailty and stratified models are step functions, so their hazards were estimated by smoothing the Breslow increments with the Epanechnikov kernel \citep{ramlau1983smoothing}, using the local linear equivalent kernel within one bandwidth of 0 and 27 days to correct the bias near the boundaries. The bandwidth was chosen from 0.5, 1, 2, 3, 4, and 6 days by the validation log-likelihood of the smoothed hazard given the estimated frailties.

\subsection{Simulation Results}

\begin{figure}[tbp]
\centering
\includegraphics[width=\linewidth]{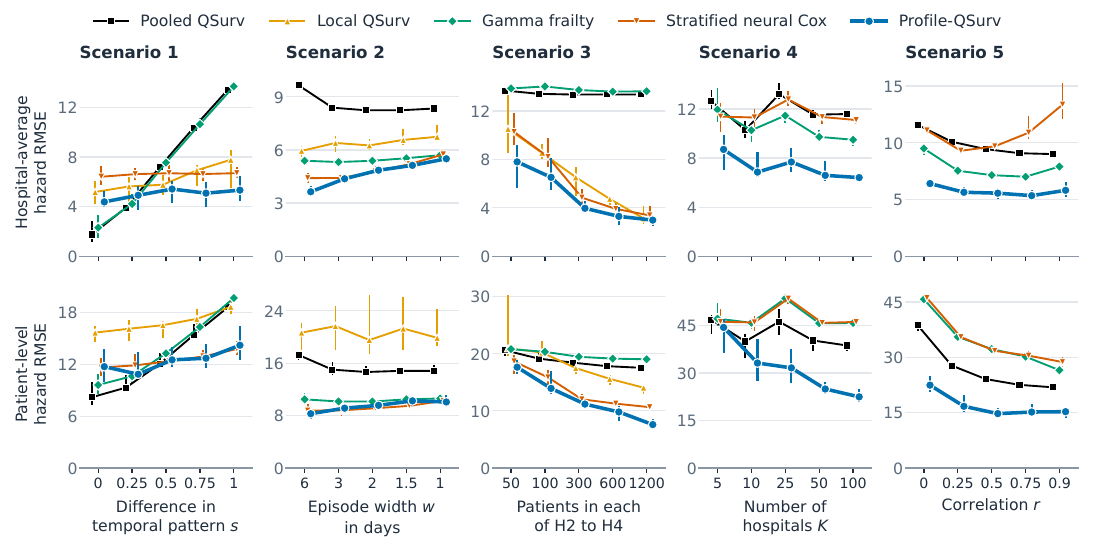}
\caption{Errors of the hospital-average hazards (top) and the patient-level hazards (bottom) in the five scenarios, in events per 1,000 patient-days. Points and bars are the medians and interquartile ranges over 20 replicates. The local model lies above the displayed range in Scenarios 4 and 5.}
\label{fig:sim}
\end{figure}

Figure \ref{fig:sim} displays the errors of the hospital-average and patient-level hazards for the five scenarios. For Scenario 1, the errors of the pooled and frailty models increased with $s$, as neither model allows the hospital differences to change over time. In contrast, the errors of \ourmethod increased little with $s$. \ourmethod had smaller hospital-average errors than the local and stratified models at every $s$, and the smallest hospital-average errors of the five methods from $s=0.5$. For the patient-level hazards, \ourmethod and the stratified model had similar errors, which were the smallest from $s=0.75$. With $s=0$, where the pooled model is correctly specified, the pooled and frailty models had the smallest errors. This is expected, since \ourmethod estimates the level of each hospital from its own events. Similarly, in the setting with proportional hospital effects, \ourmethod had slightly larger errors than the frailty model and smaller errors than the pooled and local models.

For Scenario 2, \ourmethod had the smallest errors at $w=6$, where three temporal features describe the episodes. As the episodes narrowed, the errors of \ourmethod, the frailty model, and the stratified model became closer, and \ourmethod remained the best or close to the best method at $w=2$ and $w=1$, where a quarter and a half of the variation of the episodes lie outside three functions of time.

For Scenario 3, the errors of \ourmethod, the local model, and the stratified model decreased as H2 to H4 had more deaths, while the hospital-average errors of the pooled and frailty models did not change. \ourmethod had the smallest hospital-average errors of the five methods at every size, and its patient-level errors were the smallest or close to the smallest. Its hospital-average errors approached those of the local model at 1,200 patients, while its patient-level errors remained much smaller, as its patient-risk network is learned from all hospitals. For the hospital-average hazards, the gain of \ourmethod over the local model came from the small hospitals, H2 to H4, until they reached 1,200 patients, and the two methods had similar errors in the large hospitals, H1 and H5. This is consistent with Corollary \ref{cor:attenuation}: a hospital with many events keeps its own temporal pattern, while a hospital with few events follows the temporal pattern learned from all hospitals.

For Scenario 4, \ourmethod had the smallest errors for every number of hospitals, and its patient-level errors decreased as the number of hospitals increased, as the shared networks are learned from more hospitals, while those of the other methods changed little. The local model had the largest errors, as each of its networks is trained on about 100 patients.

\begin{figure}[tbp]
\centering
\includegraphics[width=\linewidth]{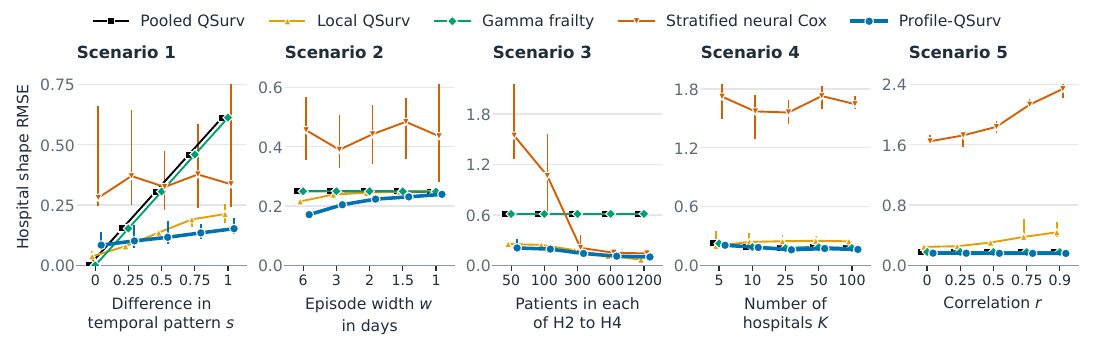}
\caption{Errors of the hospital shapes in the five scenarios. The hospital shape is the log hospital-average hazard after removing its mean over the hospitals at each time and its average over follow-up in each hospital. Points and bars are the medians and interquartile ranges over 20 replicates.}
\label{fig:sim-shapes}
\end{figure}

For Scenario 5, where the case mix of a hospital depends on its level, \ourmethod had the smallest errors at every $r$. Its patient-level errors were about half of those of the frailty and stratified models. With about 100 patients in each hospital, the separate baseline hazards of the stratified model were imprecise, and the errors of its hospital shapes, the temporal patterns of the hospitals after removing their levels, were much larger than those of \ourmethod (Figure \ref{fig:sim-shapes}).

With the penalty selected by validation, the coverage of the pointwise 95\% intervals for $\log\{c_k(t)/c_0(t)\}$ was close to 90\% in Scenarios 1 and 5, and it was lower in the datasets of Scenario 1 in which validation selected the largest penalty while the hospitals differed in their temporal patterns. With the hospital parameters profiled at $\rho=1$ on the same networks, the coverage was closer to the nominal level in these datasets and in Scenario 5. The Monte Carlo standard errors of the coverages in Scenario 1 are at most 4 percentage points.

\section{COVID-19 Hospital Comparison Case Study}
\label{sec:application}

We applied \ourmethod, in comparison with other methods, to adult COVID-19 admissions in the hospitals of S\~{a}o Paulo, Brazil, in 2020. The goals were to describe how the hospital differences in the death hazard change over the stay, to identify the hospitals whose death hazard is above that of a typical hospital and the part of the stay in which it is, to determine which deaths of these hospitals to review, and to separate differences in death from differences in discharge in the comparison of in-hospital mortality.

\subsection{Data}

SIVEP-Gripe is the national surveillance system for severe acute respiratory illness in Brazil, in which the notification of hospitalized cases is mandatory \citep{ranzani2021characterisation}. From the public release of the 2020 records \citep{sivep2026}, we identified 73,019 admissions of patients aged 18 years or older with COVID-19 as the final classification, admitted between March 1 and December 31, 2020, to a facility in the city of S\~{a}o Paulo. Each record names one hospital of admission, to which we assigned the admission and its outcome. We determined establishment types by matching the admitting facilities to the National Registry of Health Establishments (CNES), with manual review. We successively excluded 1,652 hospital-acquired infections, 913 admissions with an unknown or inconsistent outcome, 5,435 admissions to facilities other than general and specialized hospitals, such as field hospitals and emergency care units, and 963 admissions to hospitals with fewer than 50 admissions or 10 deaths within 28 days, so that the level of every hospital is estimable. Table \ref{tab:app-characteristics} describes the remaining 64,056 admissions to 106 hospitals. The hospitals admitted between 50 and 3,982 patients, and their crude probability of death in hospital by day 28 ranged from 2.6\% to 80.6\%.

\begin{table}[tbp]
\centering
\caption{Characteristics at admission and outcomes by day 28. Values are counts (percentages) or medians (interquartile ranges), and comorbidities and signs are counted when recorded as present.}
\label{tab:app-characteristics}
\begin{tabular}{lr}
\hline
 & Cohort \\
\hline
Admissions & 64,056 \\
Age, years & 60 (47 to 73) \\
Female & 28,318 (44.2) \\
Race or skin color, white & 27,792 (43.4) \\
Race or skin color, brown & 12,802 (20.0) \\
Race or skin color, black & 3,055 (4.8) \\
Race or skin color, asian & 1,183 (1.8) \\
Race or skin color, indigenous & 69 (0.1) \\
Race or skin color, unknown & 19,155 (29.9) \\
Heart disease & 25,585 (39.9) \\
Diabetes & 16,935 (26.4) \\
Chronic lung disease & 2,912 (4.5) \\
Chronic kidney disease & 3,033 (4.7) \\
Obesity & 4,316 (6.7) \\
Immunosuppression & 2,391 (3.7) \\
Neurological disease & 3,324 (5.2) \\
Asthma & 2,012 (3.1) \\
Dyspnea & 45,465 (71.0) \\
Respiratory distress & 34,790 (54.3) \\
Oxygen saturation below 95\% & 38,982 (60.9) \\
Days from symptom onset to admission & 6 (3 to 9) \\
Residence outside the city & 12,336 (19.3) \\
Relative load at admission & 0.66 (0.50 to 0.83) \\
Death in hospital by day 28 & 14,468 (22.6) \\
Discharge alive by day 28 & 43,780 (68.3) \\
In hospital at day 28 & 5,808 (9.1) \\
\hline
\end{tabular}
\end{table}

We modeled the hazards of in-hospital death from any cause and of discharge alive over the first $\tau=28$ days from admission. Admissions that ended after day 28 were censored at day 28, and outcomes on the day of admission were placed at 0.5 days. We fitted model \eqref{eq:hazard-model} separately to the two outcomes, treating the other outcome as censoring. We write
\begin{equation}
\lambda_{kj}(t|x)=c_{kj}(t)h_j(t,x), j=1,2,
\label{eq:app-model}
\end{equation}
for the hazard of death ($j=1$) and of discharge alive ($j=2$) in hospital $k$. The death hazard among patients still in hospital is the primary target, and the absolute risk summaries below use both hazards.

The patient-risk networks received the information recorded at admission: age, sex, race or skin color, eight comorbidities and three signs of severity at presentation (Table \ref{tab:app-characteristics}), the days from symptom onset to admission, the admission date, residence outside the city, and the relative load of the admitting hospital. Comorbidities and signs were coded as present, absent, or unknown, so that admissions with incomplete records remain in the cohort. Race or skin color was associated with the in-hospital mortality of COVID-19 in SIVEP-Gripe \citep{baqui2020ethnic}, and surges in caseload were associated with higher COVID-19 mortality \citep{kadri2021association,brizzi2022spatial}. The relative load is the number of inpatients with severe acute respiratory illness in the hospital on the day of admission, counted from all of its SIVEP-Gripe records with admission and outcome dates and divided by its maximum over 2020. With this covariate, the hospital factors compare hospitals at the same relative load. Treatments and admission to intensive care occur after admission and belong to the care that the hospital factors compare.

\subsection{Analysis}

Every model was tuned as in the simulations (Section \ref{sec:simulation}). For each outcome, each model had ten network configurations with SiLU activations, sampled from the same combinations of widths, numbers of hidden layers, learning rates, and weight decay, and each configuration of \ourmethod was paired with one penalty $\rho\in\{1,10,100\}$. The temporal network had one hidden layer of eight units with $q=3$ features, and the exposure integrals used 12 Gauss-Legendre nodes on each of the intervals $[0,3]$, $[3,7]$, $[7,14]$, and $[14,28]$ days. The five folds were formed within hospitals. In each fold, a model was trained for 300 epochs on three parts, the epoch with the largest validation log-likelihood on a fourth part was kept, with the hospitals weighted equally, and the fifth part was held out. For each model, the configuration with the largest validation log-likelihood averaged over the five folds was selected and, for inference, trained for the median of its selected numbers of epochs. The baseline hazards of the stratified neural Cox model were smoothed as those of the frailty model in Section \ref{sec:simulation}, with the bandwidth chosen by the validation log-likelihood in each fold.

We compared the death hazard of \ourmethod with the pooled QSurv, local QSurv, neural gamma frailty, and stratified neural Cox models of Section \ref{sec:simulation}. We divided the stay into three phases, $\mathcal{I}_1=[0,3)$, $\mathcal{I}_2=[3,14)$, and $\mathcal{I}_3=[14,28]$ days, which follow the clinical course described in Section \ref{sec:introduction}: the first days after admission, the rest of the first two weeks, and the period after the second week. The methods were compared by the held-out negative log-likelihood on daily intervals and by a phase calibration statistic. The phase statistic is the mean of $(O_{km}-E_{km})^2/E_{km}$ over hospitals and phases, where $O_{km}$ and $E_{km}$ are the observed and expected deaths of hospital $k$ in phase $m$, and it was computed in each held-out fold and averaged over the five folds. An outcome recorded on day $j$ belongs to the daily interval $(j-1,j]$, so the phase statistic compares the deaths recorded up to day 3, on days 4 to 14, and on days 15 to 28 with the expected deaths over the three phases. The phase statistic of each fold is taken over the hospitals and phases with a death in the training parts of the fold. Both criteria use, in each fold, the configuration with the largest validation log-likelihood in that fold. By Proposition \ref{prop:calibration}, the observed and expected deaths of each hospital agree over the whole follow-up in the training data. The phase statistic examines this agreement within the phases of the stay on held-out admissions. To compare \ourmethod with each other method where their predictions differ, we counted the held-out deaths of each fold by hospital and phase, grouped the counts into tenths by the ratio of the deaths expected under \ourmethod to those expected under the other method, and compared the observed deaths of each group with the two expected numbers. The groups were formed within the folds, so that the grouping and the expected deaths of each count come from models trained without its deaths. We also fitted a Poisson regression of the held-out deaths on the logarithm of this ratio, with the logarithm of the deaths expected under the other method as offset and standard errors clustered by hospital. Its coefficient is one when the deaths follow \ourmethod and zero when they follow the other method.

For inference, the shared networks were trained on a random half of the admissions of each hospital, and the hospital parameters were profiled on the other half with the networks fixed. The halves were formed separately within each hospital for deaths, discharges, and admissions still in hospital at day 28, so that each half received half of the deaths and half of the discharges of the hospital, and at least five deaths. With fixed networks, the parameters $(\log a_k,w_k)$ maximize a penalized likelihood whose information follows from \eqref{eq:profile-hessian}. Its inverse is the approximate posterior covariance under the Gaussian prior implied by the penalty, and gives
\begin{equation}
\text{\normalfont Var}\{\log\widehat{c}_k(t)\}=\frac{1}{D_k}+\{\phi(t)-m_k\}^{T}\{D_k\text{\normalfont Cov}_k(\phi)+\rho I_q\}^{-1}\{\phi(t)-m_k\},
\label{eq:app-variance}
\end{equation}
where $D_k$ is the number of events of hospital $k$ in the profiling half, and $m_k$ and $\text{\normalfont Cov}_k(\phi)$ are the moments of the temporal features under its fitted exposure distribution. The roles of the halves were then swapped, and the hospital factors and the summaries derived from them average the two. The two halves also give two estimates of the factors of each hospital from disjoint sets of its admissions, which we compare to examine whether the changes over the stay are reproducible. Pointwise 95\% intervals come from 200 draws of the hospital parameters of each half from their approximate normal distributions with the networks fixed, independent across hospitals and halves. The coverage of these intervals is examined in Section \ref{sec:simulation}.

The hospital factors are identified up to a common function of time, so we compare each hospital with a reference hospital. The factors of the reference hospital are the pointwise medians $c_{0j}(t)=\text{\normalfont median}_kc_{kj}(t)$, as in time-dynamic profiling \citep{estes2018time}. For each hospital $k$ and phase $m$, let $\bar{\ell}_{km}$ denote the average of $\log\{c_{k1}(t)/c_{01}(t)\}$ over a grid of the phase with steps of 0.25 days, with its 95\% interval formed by the 2.5\% and 97.5\% quantiles of the same average over the draws. We identify the hospital as above the reference in the phase when the 95\% interval of $\bar{\ell}_{km}$ lies above zero, as providers that differ from a reference are identified in profiling \citep{kalbfleisch2013monitoring}. In time-dynamic profiling, facilities are classified as consistently above the reference, consistently below it, or mixed over follow-up \citep{estes2018time}.

The phases also indicate which deaths of a hospital to review. Hospitals examine the care of patients who died through reviews of case records, and a hospital selects the deaths that it reviews \citep{kobewka2017quality}. At time $t$, a fraction $1-c_{01}(t)/c_{k1}(t)$ of the death hazard of a hospital above the reference is in excess of the reference. We summarize it in phase $m$ by
\begin{equation}
\pi_{km}=\max\{0,1-\exp(-\bar{\ell}_{km})\},
\label{eq:app-share}
\end{equation}
the expected share of the deaths of hospital $k$ in phase $m$ that are in excess of the reference. Let $D_{km}$ be the number of deaths of hospital $k$ in phase $m$, counted as for the phase statistic. A hospital that reviews a sample of its deaths taken over the whole stay expects a share $\sum_mD_{km}\pi_{km}/\sum_mD_{km}$ of excess deaths among them. We compare it with the share obtained when the hospital takes the deaths from the phases in decreasing order of $\pi_{km}$. To evaluate the order on admissions that did not determine it, the order was taken from one half of the admissions and the shares $\pi_{km}$ from the other half, and the two directions were averaged. Under a hospital effect that is constant over the stay, $\pi_{km}$ is the same in every phase, and every order gives the share of the sample over the whole stay.

To separate death from discharge, we use the probability of death in hospital by day $\tau$ for covariates $x$ and factors $(c_1,c_2)$,
\begin{equation}
F(x;c_1,c_2)=\int_0^\tau c_1(t)h_1(t,x)\exp\left[-\int_0^t\{c_1(u)h_1(u,x)+c_2(u)h_2(u,x)\}du\right]dt.
\label{eq:app-cif}
\end{equation}
We computed $F$ by the trapezoidal rule on a grid with steps of 0.025 days over the first day and 0.1 days afterward. Let $p(c_1,c_2)$ denote the average of $F(x;c_1,c_2)$ over all admissions, the standardized probability of death under factors $(c_1,c_2)$, so that $p_k=p(c_{k1},c_{k2})$ for hospital $k$ and $p_0=p(c_{01},c_{02})$ for the reference. Here and below, averages of $F$ over more than 3,000 admissions were computed on a fixed random sample of 3,000 of them. The difference $p_k-p_0$ splits into the death component
\begin{equation}
\Delta_{k1}=\frac{1}{2}\left[\{p(c_{k1},c_{02})-p(c_{01},c_{02})\}+\{p(c_{k1},c_{k2})-p(c_{01},c_{k2})\}\right]
\label{eq:app-decomposition}
\end{equation}
and the discharge component $\Delta_{k2}=p_k-p_0-\Delta_{k1}$, averaged over the two orders of replacing the factors. A hospital with a negative discharge component records fewer deaths in hospital because its patients leave the hospital sooner than at the reference.

We also carried out two further analyses, of the association of the hazards with the load of the hospital and of the order of the hospitals for two groups of patients. The relative load enters the shared networks. Because the hospital factors absorb the differences between hospitals that persist over the year, the association of load with the hazards is learned from the variation of load within hospitals, and the admission date absorbs the changes common to all hospitals over the epidemic, so the association is identified by the hospitals reaching their largest census at different dates. Let $\bar{x}_{ik}$ be the covariates of patient $i$ in hospital $k$ with the relative load capped at $1/2$, and let $\mathcal{A}$ be the set of admissions with relative load above $1/2$. For outcome $j$, we summarize the association by
\begin{equation}
\text{\normalfont HR}_j(t)=\exp\left[\frac{1}{|\mathcal{A}|}\sum_{(i,k)\in\mathcal{A}}\left\{\log h_j(t,x_{ik})-\log h_j(t,\bar{x}_{ik})\right\}\right],
\label{eq:app-load}
\end{equation}
which compares, for patients admitted when their hospital was above half of its largest census, the hazard on day $t$ at the observed load with the hazard at half of the largest census. The hospital factors cancel in this ratio. We report these hazard ratios averaged over the two halves of the sample split and from each half.

The hospital factors are shared by all patients of a hospital, but patients differ in when their risk is concentrated, so the order of hospitals by probability of death can depend on the patients being compared. For a group $\mathcal{G}$ of admissions, the standardized probability of death in hospital $k$ is
\begin{equation}
p_k(\mathcal{G})=\frac{1}{|\mathcal{G}|}\sum_{i\in\mathcal{G}}F(x_i;c_{k1},c_{k2}),
\label{eq:app-standardized}
\end{equation}
the probability of death if the patients in $\mathcal{G}$ were admitted to hospital $k$. We compared two groups from the whole cohort: patients younger than 60 years without recorded comorbidities or signs of severity, and patients aged 60 years or older with oxygen saturation below 95\% or respiratory distress. The agreement of the two orders was summarized by the rank correlation of $p_k(\mathcal{G})$ across hospitals and by the overlap of the hospitals in the lowest quarter for the two groups.

\subsection{Results}

Cross-fitting selected the penalty $\rho=100$ for both the death and the discharge hazards. \ourmethod had the smallest held-out negative log-likelihood and phase calibration statistic of the five methods (Table \ref{tab:app-comparison}). The pooled model, which assigns the same hazard to patients with the same covariates in every hospital, had the largest phase statistic, and the pooled and local models had the largest negative log-likelihoods.

\begin{figure}[t]
\centering
\includegraphics[width=0.99\linewidth]{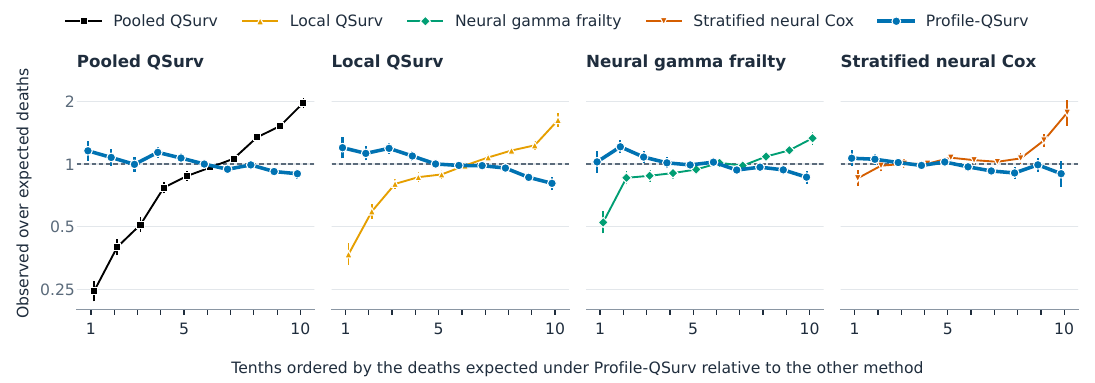}
\caption{Observed over expected held-out deaths under \ourmethod and each other method. The held-out deaths of each fold were counted by hospital and phase, and for each other method the counts were grouped into tenths by the ratio of the deaths expected under \ourmethod to those expected under that method. Each panel compares \ourmethod with one method on the tenths of that comparison. Vertical bars are 95\% intervals from the Poisson variance of the observed deaths, and the vertical axis is on a log scale.}
\label{fig:app-heldout}
\end{figure}

Figure \ref{fig:app-heldout} compares the observed held-out deaths with the deaths expected under \ourmethod and under each other method. In the first and the last tenth, where the two methods differed the most, the observed deaths were closer to those expected under \ourmethod in all four comparisons. In the tenth of the counts in which \ourmethod expected the fewest deaths relative to the frailty model, 262 deaths were observed, against 256 expected under \ourmethod and 500 under the frailty model, and in the tenth in which it expected the most, 747 deaths were observed, against 865 and 562. The coefficient of the Poisson regression was 0.88 against the pooled model, 0.73 against the local model, 0.70 against the frailty model, and 0.80 against the stratified model, with standard errors between 0.02 and 0.10. Against the frailty model, it was 0.92 in days 0 to 2, 0.41 in days 3 to 13, and 0.76 in days 14 to 28, so the held-out deaths were closer to \ourmethod in the first days and after the second week, the phases in which a hospital effect that is constant over the stay departs from the hospital factors.

\begin{table}[t]
\centering
\caption{Held-out fit of the death hazard under fivefold cross-fitting within hospitals. NLL is the negative log-likelihood per admission on daily intervals, and the phase calibration statistic is the mean of $(O_{km}-E_{km})^2/E_{km}$ over hospitals and the three phases.}
\label{tab:app-comparison}
\begin{tabular}{lcc}
\hline
Model & NLL & Phase calibration \\
\hline
Pooled QSurv & 1.0341 & 3.08 \\
Local QSurv & 1.0336 & 1.87 \\
Neural gamma frailty & 1.0090 & 1.47 \\
Stratified neural Cox & 1.0142 & 1.58 \\
Profile-QSurv & 1.0071 & 1.36 \\
\hline
\end{tabular}
\end{table}

\begin{figure}[t]
\centering
\includegraphics[width=0.99\linewidth]{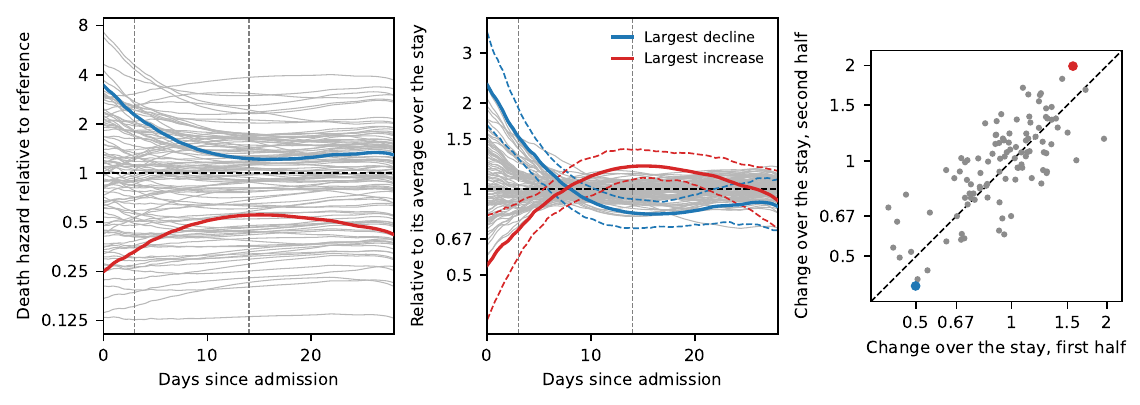}
\caption{Death hazard of each hospital relative to the reference hospital, $c_{k1}(t)/c_{01}(t)$ (left), the same ratio divided by its geometric mean over the stay (middle), and the change of the ratio from days 0 to 2 to days 14 to 28 estimated from each of the two halves of the admissions (right). All axes are on a log scale. The highlighted hospitals have the largest decline and the largest increase from days 0 to 2 to days 14 to 28, with pointwise 95\% intervals in dashed lines.}
\label{fig:app-death}
\end{figure}

Figure \ref{fig:app-death} displays the estimated death factors relative to the reference, $c_{k1}(t)/c_{01}(t)$. The differences between hospitals in the death hazard were largest at admission, when differences in severity at presentation and in the escalation of care are expected to appear \citep{churpek2016association}, and narrowed over the stay. Across hospitals, the ratio of the 90th to the 10th percentile of the death factor averaged over each phase was 9.0 in days 0 to 2, 6.5 in days 3 to 13, and 4.9 in days 14 to 28. Within a hospital, the death hazard relative to the reference also varied over the stay: the ratio of its largest to its smallest value had a median of 1.36 across hospitals and exceeded 1.5 for 35 of the 106 hospitals. For 79 hospitals, it was farthest from its average over the stay in the first three days. The order of the hospitals was stable over the stay, with a rank correlation of 0.93 between the death factors of days 0 to 2 and of days 14 to 28, so the hospital differences persisted and changed mainly in size. The changes over the stay were reproducible between the two halves of the admissions. The change of the death factor of a hospital from days 0 to 2 to days 14 to 28 had a correlation of 0.74 between the halves (right panel of Figure \ref{fig:app-death}), and it had the same direction in both halves for each of the 48 hospitals in which it was larger than 1.25-fold in either direction.

Forty-three hospitals had a death hazard above the reference in at least one phase. Thirty of them were above the reference in all three phases. The other 13 were above it in part of the stay: 7 from day 3 onward, 4 in days 0 to 2 or days 0 to 13, and 2 in days 3 to 13. Among the 43 hospitals, the death hazard relative to the reference was highest in days 0 to 2 for 28, and the phase with the highest value was the same in the two halves of the admissions for 32. A death hazard above the reference in the first days and after the second week may reflect different parts of the care \citep{churpek2016association,contou2021causes}. For most of the identified hospitals, the difference from the reference was largest in the first days after admission, which indicates this part of the care as the first to review. For the 7 hospitals that were above the reference from day 3 onward, the intervals included the reference in the first days, so their review concerns the later care.

For the same 43 hospitals, the median share of the deaths in excess of the reference, $\pi_{km}$, was 52\% in days 0 to 2, 40\% in days 3 to 13, and 36\% in days 14 to 28, and the largest and smallest shares of a hospital differed by more than 15 percentage points in 25 hospitals. When each hospital reviewed one fifth of its deaths, taken from the phases in decreasing order of $\pi_{km}$, 50\% of the reviewed deaths were in excess of the reference on average over the hospitals, compared with 42\% for a sample over the whole stay and 51\% for the best order. The share increased by more than 5 percentage points in 27 hospitals, by up to 22 points, and decreased by more than 5 points in none. The results were similar when one tenth or three tenths of the deaths were reviewed, with 50\% and 49\% of the reviewed deaths in excess of the reference. The frailty model assigns each hospital a single effect for the whole stay, so the share is the same in every phase and every order of the deaths is equivalent, and the pooled model has no hospital effect. The stratified and local models estimate the hazard of each hospital from its own patients, which was less accurate than \ourmethod in the simulations with small hospitals (Section \ref{sec:simulation}) and on the held-out deaths here (Table \ref{tab:app-comparison} and Figure \ref{fig:app-heldout}).

The comparison of in-hospital mortality also depended on discharge. The discharge component ranged from $-6.1$ to 25.4 percentage points across hospitals. Of the 48 hospitals with standardized in-hospital mortality below the reference, 4 had a death component above zero and a discharge component below zero. Their in-hospital mortality was below the reference because their patients were discharged sooner. This is consistent with \citet{drye2012comparison}, who found that in-hospital mortality favors hospitals with short stays when it is compared with 30-day mortality.

For the 47,386 admissions above half of the largest census of their hospital, discharge was faster at the observed load than at half of the census, with hazard ratios of 1.05 on day 7 and 1.06 on day 21, and between 1.03 and 1.07 in each half of the sample split. The hazard ratio of death on day 1 was 1.01 in one half and 0.90 in the other. The order of the hospitals was similar for the two groups of patients, with a rank correlation of 0.95 (95\% interval 0.93 to 0.96) between their standardized probabilities of death, and 25 of the 27 hospitals in the lowest quarter for the older patients with hypoxemia or respiratory distress were also in the lowest quarter for the younger patients without comorbidities or signs.

\section{Conclusions and Discussion}
\label{sec:discussion}

In this study we introduced \ourmethod for estimating cluster-specific hazards in multicenter survival data. A shared neural network learns how covariates and time relate to the hazard, and each cluster carries an unpenalized level and a penalized shape on a shared temporal dictionary. The level is profiled in closed form and the shape through a small convex problem. Further, we showed that each cluster remains calibrated on its own events, and that a change in the shared hazard is divided between the fitted hazard and the cluster parameters. The division is governed by a shrinkage factor determined by the event count of the cluster and the penalty.

The profile structure also gives a direct route to inference for cluster trajectories. The cluster hazard ratio \eqref{eq:cluster-ratio} depends only on the cluster parameters and the dictionary. Given shared networks fitted on an independent part of the data, these parameters solve a finite-dimensional penalized likelihood whose observed information gives pointwise intervals, as in Section \ref{sec:application}. These intervals condition on the trained networks. We examined their coverage in the simulations of Section \ref{sec:simulation}. The effects of estimating the shared networks and of the shrinkage on the intervals need to be further investigated.

Moving forward, \ourmethod can be extended in two directions. Prediction for a new cluster requires a distribution for the cluster parameters. This distribution can be estimated from the fitted levels and shapes after accounting for their estimation error and shrinkage, which connects the method with marginal prediction from frailty models. Cluster-specific covariate effects can be added through low-rank cluster terms in the patient-risk network, with penalties that play the same role as $\rho$.


\bibliographystyle{abbrvnat}  
\bibliography{references}

\end{document}